 \documentclass[twocolumn]{aastex631}

\shorttitle{Light curve model of CN Cha}
\shortauthors{Kato \& Hachisu}

\begin{document}

\title{Theoretical light curve models of the symbiotic nova CN Cha ---
Optical flat peak for three years
}


\author[0000-0002-8522-8033]{Mariko Kato}
\affil{Department of Astronomy, Keio University, 
Hiyoshi, Kouhoku-ku, Yokohama 223-8521, Japan} 
\email{mariko.kato@hc.st.keio.ac.jp}



\author[0000-0002-0884-7404]{Izumi Hachisu}
\affil{Department of Earth Science and Astronomy, 
College of Arts and Sciences, The University of Tokyo,
3-8-1 Komaba, Meguro-ku, Tokyo 153-8902, Japan} 



\begin{abstract}
CN Cha is a slow symbiotic nova characterized by a three-years-long
optical flat peak followed by a rapid decline.  We present theoretical
light curves for CN Cha, based on hydrostatic approximation,
and estimate the white dwarf (WD) mass to be $\sim 0.6~M_\sun$
for a low metal abundance of $Z=0.004$. 
This kind of flat peak novae are border objects
between classical novae having a sharp optical peak 
and extremely slow novae, the evolutions of which are too slow 
to be recognized as a nova outburst in human timescale.  
Theoretically, there are two types of nova envelope solutions,
static and optically-thick wind,
in low mass WDs ($\lesssim 0.7 ~M_\sun$).
Such a nova outburst begins first in a hydrostatic manner,
and later it could change to an optically-thick wind evolution
due to perturbation
by the companion star in the nova envelope.  Multiple peaks are a
reflection of the relaxation process of transition.
CN Cha supports our explanation on the difference between long-lasted
flat peak novae like CN Cha and multiple peak novae like V723 Cas, 
because the companion star is located far outside, and does not perturb,
the nova envelope  in CN Cha.
\end{abstract}




\section{Introduction} \label{introduction}

CN Cha is a galactic symbiotic nova outbursted in late 2012
or early 2013. It was identified as a Mira variable \citep{hof63}
long before the outburst.  The outburst shows a stable flat peak
at $m_V \sim 8$ mag that lasted three years
followed by a rapid decline.   \citet{lan20} summarized observational 
information in literature and database and made a comprehensive $V$ 
light curve (see their Figure 2).
They also presented an optical/near IR spectrum taken on UT 2019 
March 12 that shows emission lines including P-Cygni profiles.
The light curve of CN Cha resembles the eight-years-long flat peak of
the symbiotic nova PU Vul.

CN Cha is located one kpc below the galactic plane for the Gaia early
Data Release 3 (Gaia eDR3) distance of 
$d= 3.05^{+0.19}_{-0.17}$ kpc \citep{bai21rf}. 
\citet{lan20} presented a possible historical orbit in 
the Galaxy and concluded that this star is most likely 
a thick disk component star of $> 8$ Gyr old. 
Note that wind mass loss is systematically weaker 
and their evolutions are slower in Population II novae \citep{kat13hh}
than in Population I novae \citep[e.g.,][for a summary]{del02}.


\begin{figure*}
\epsscale{0.85}
\plotone{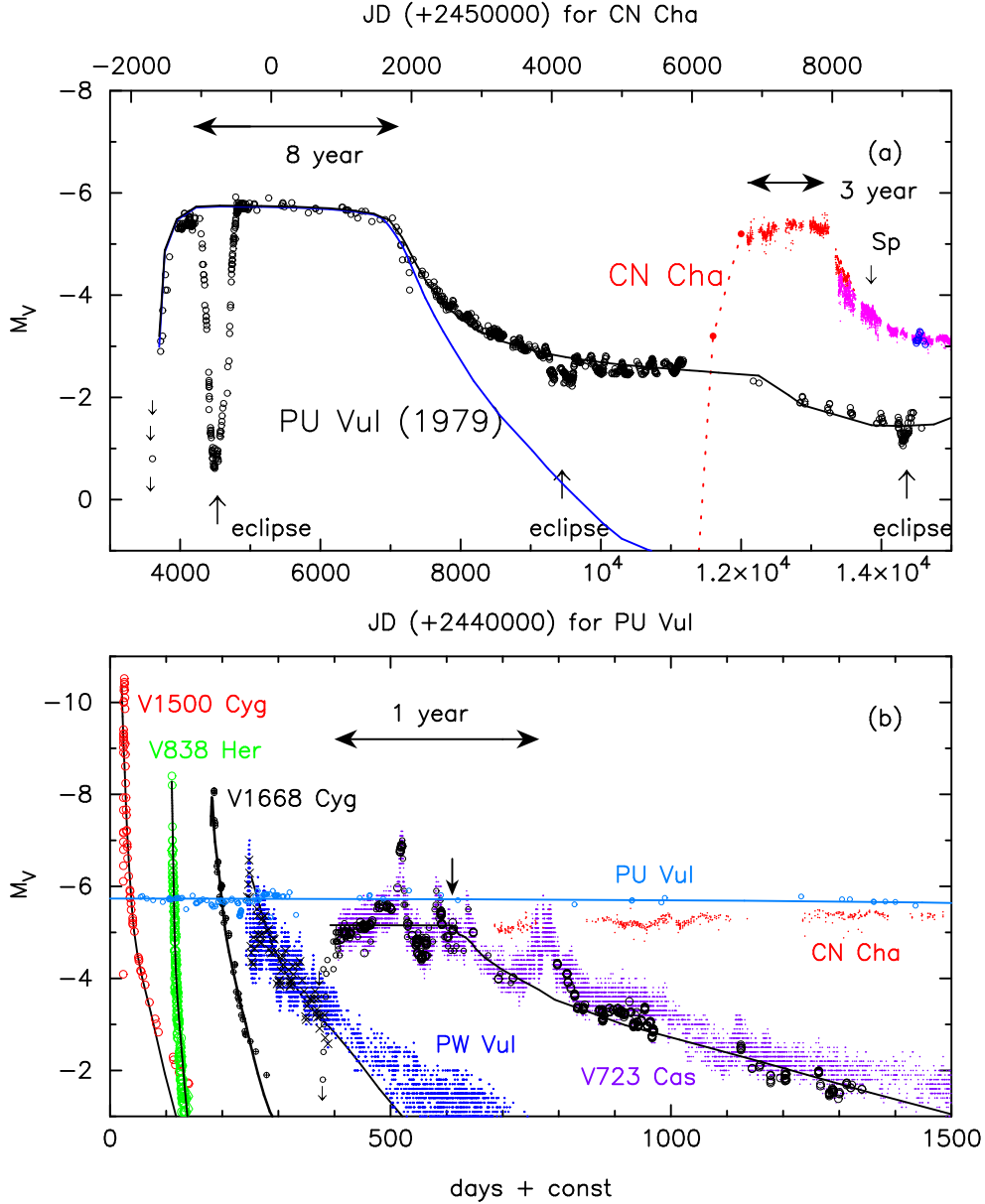}
\caption{Comparison of optical light curves of CN Cha with 
various types of novae. 
Data of each nova other than CN Cha are the same as those in Figure 1 of
\citet{kat13hh}. 
Time and absolute $V$ magnitude, $M_V$,
are plotted in the same scale for all the objects in each panel. 
The solid lines represent theoretical models for each object. 
See the main text for more detail.  
(a) The very slow nova PU Vul 1979 together with CN Cha.  We adopt the
distance modulus in the $V$ band of $\mu_V\equiv (m-M)_V=14.3$ for PU Vul,
where the distance $D=4.7$ kpc and extinction $E(B-V)=0.3$ are taken from 
\citet{kat12}. The black line denotes the model light curve
of PU Vul which is the summation of the photospheric blackbody emission 
(blue line), and emission from optically thin plasma \citep{kat12}. 
The upward arrows indicate the mid eclipses in PU Vul.
For CN Cha we assume $\mu_V\equiv (m-M)_V=13.2$.
(b) The very fast novae V1500 Cyg and V838 Her, fast nova V1668 Cyg,  
moderately fast nova PW Vul, and slow nova V723 Cas. 
The flat peak in PU Vul and CN Cha partly appears in this timescale. 
The solid line in V723 Cas represent a binary model consisting of a
0.6 $M_\sun$ WD and 0.4 $M_\sun$ main-sequence companion \citep{kat11drag};  
the downward arrow indicates the transition point from static 
to wind evolution. 
\label{light.compar}}
\end{figure*}

A long-lasted flat-peak rarely appears among a number of novae,
whereas many show a sharp optical maximum \citep[e.g.,][]{str10}. 
A flat peak in novae is theoretically explained as follows: 

A nova is a thermonuclear runaway event on a mass-accreting white dwarf (WD) 
\citep{nar80, ibe82, pri86, pri95, sio79, spa78, kat17}.
Once hydrogen shell burning begins, a hydrogen-rich envelope atop the 
WD expands to a giant size.  Strong optically-thick winds are 
accelerated \citep{kat94h} that blow off a large part of the envelope. 
The nova evolves fast to reach its optical maximum 
and immediately enters the decay phase. 
This wind mass loss accelerates the nova evolution and, as a result,
stronger wind mass-loss makes a sharper maximum in the optical light curve
\citep[e.g.,][]{hac20skhs}.   

In general, the wind is stronger in more massive WDs 
and/or larger heavy element enrichment \citep{hac10k,kat13hh}. 
On the other hand, in less massive WDs and/or lower metallicity environments, 
the nova evolves more slowly because of their weaker mass losses. 
In an extreme case, no optically thick winds are 
accelerated and, then, the nova evolves very slowly \citep{kat09}.
The envelope structure hardly changes because its mass is decreasing
due only to hydrogen shell-burning, about a thousand times longer
timescale compared with that due to strong winds.  Therefore, the nova
stays at an expanded stage for a long time.
This makes the optical peak flat for a thousand times longer time. 


A good example of long-lasted flat peak is the symbiotic nova PU Vul
\citep[][for a recent summary]{cun18kg}. 
This is a well observed nova, but there is no indication of optically-thick
wind mass-loss \citep[e.g.,][]{yam82, kan91a}.
The flat peak lasts as long as eight years 
(see Figure \ref{light.compar}).   If such a flat peak lasts much longer,
e.g., a century, the outbursting WD may not be recognized as a nova
outburst, instead as a supergiant.  The flat-peak nova PU Vul could be
an object on the border that we can recognize a thermonuclear
runaway event as a nova in human timescale.  

Theoretically a hydrogen shell flash occurs on a WD of mass as small as
0.4 $M_\sun$ \citep{nar80, yar05, she09}. 
On the other hand, estimated WD masses in novae 
are more massive \citep[$> 0.55-0.6 M_\sun$, see, e.g.,][]{hor93ww,
smi98dm, tho01dlms, kat11drag, hac15k, sel19g}.
A possible explanation of this discrepancy is 
that we have missed such variables fading very slowly after 
red-giant-like stage.

\citet{kat09} studied the condition of occurrence of optically 
thick winds. The border of WD mass for occurrence of winds lies
at $M_{\rm WD,cr} \approx 0.6$ - 0.7 $M_\sun$, depending on the metallicity. 
A shell flash on a WD ($M_{\rm WD} < M_{\rm WD,cr}$) evolves 
too slow to be recognized as a nova outburst in human timescale.
PU Vul is a rare object that was identified as a nova close to this border. 

In this paper, we present light curve models for CN Cha 
and determine the WD mass.  We also discuss whether or not CN Cha is
the next close example of the border object.
This paper is organized as follows.
First, we compare the light curve of CN Cha with various types of novae 
in Section \ref{sec_optical}. Then, we present our model light curves
to estimate the WD mass in section \ref{section_model}. 
Discussion and conclusions follow in sections \ref{discussion}
and \ref{conclusion}, respectively.

\begin{figure}
\epsscale{1.16}
\plotone{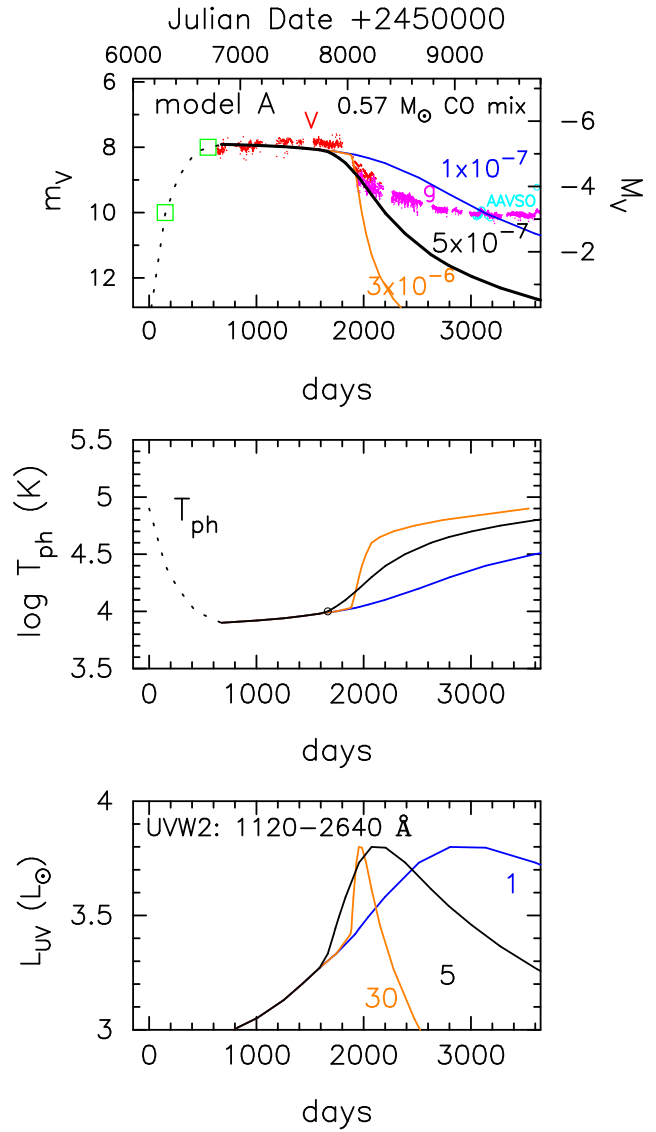}
\caption{
{\bf Top panel:} 
The theoretical $V$ light curves of model A in Table \ref{table_model}. 
Observed data are taken from ASAS-SN
($V$: red dots, $g$: magenta dots), 
AAVSO (cyan blue dots) and \citet{lan20} (green squares).  
The dotted line parts represent the pre-maximum
phase in which we determined $T_{\rm ph}$ to be consistent with 
observational estimates (green squares).
The optically thin mass loss is assumed when the photospheric
temperature rises to $\log T_{\rm ph}$ (K) $> 4.0$.
The orange, black, and blue lines correspond to the optically thin
wind mass-loss rates of ${\dot M}_{\rm wind}= 3\times 10^{-6}$,
$5 \times 10^{-7}$, and $1\times 10^{-7}~M_\sun$ yr$^{-1}$, respectively. 
{\bf Middle panel:} 
The photospheric temperature  $\log T_{\rm ph}$ for each model. 
The open circle indicates the epoch when optically thin
mass loss begins. 
{\bf Bottom panel:} 
The temporal change of UV flux (UVW2: 1120-2640 \AA)
for each $\dot{M}_{\rm wind}$ model.
 The wind mass loss rate is attached beside the curve in units 
of $10^{-7} ~M_\sun$ yr$^{-1}$. 
\label{m057.CO}}
\end{figure}

\begin{deluxetable*}{clccccccccccc}
\tabletypesize{\scriptsize}
\tablecaption{Model parameters
\label{table_model}}
\tablewidth{0pt}
\tablehead{\colhead{Model }& \colhead{} &
\colhead{$M_{\rm WD}$}&
\colhead{$X$}& \colhead{$Y$} & \colhead{$Z$} & \colhead{$Z_{\rm CO}$} &
 \colhead{$\log R_{\rm WD}$} & 
\colhead{$M_{\rm ig}$} &\colhead{$(m-M)_V$}&\colhead{$\dot{M}_{\rm wind}$} \\
 \colhead{} & \colhead{} & \colhead{($M_\sun$)} &\colhead{}& \colhead{} & 
 \colhead{} & \colhead{} & \colhead{($R_\sun$)}  & \colhead{($10^{-5}~M_\sun$)}&
 \colhead{} &  \colhead{($10^{-7}~M_\sun$ yr$^{-1}$)}
}
\startdata
A & ... &0.57& 0.70 &0.296 & 0.004 & 0.2 &-1.82 &$4.4$ & 13.2 &  $1,5,30$   \\
B & ... &0.55& 0.70 &0.296 & 0.004 & 0.2&-1.80& $5.5$ & 13.05 & $2,5,20$  \\
C & ... &0.6& 0.70 &0.296 & 0.004 & 0.0 &-1.83& $14$ & 13.05 & $5,50,80$\\
D & ... &0.6& 0.70 &0.29 & 0.01 & 0.0 &-1.90& $4.0$ & 13.2 & $5,30$\\
\enddata
\end{deluxetable*}

\section{Comparison of Optical Light Curves among Different Speed Classes}
\label{sec_optical}

Figure \ref{light.compar} shows light curves of well observed novae 
having different speed classes\footnote{The nova speed class is defined
by $t_3$ or $t_2$ (days of 3 or 2 mag decay from optical maximum).
For example,
very fast novae ($t_2 \le 10$ day),
fast novae ($11 \le t_2 \le 25$ day),
moderately fast nova ($26 \le t_2 \le 80$ day),
slow novae ($81 \le t_2 \le 150$ day),
and very slow novae ($151 \le t_2 \le 250$ day),
which are defined by \citet{pay57}.}.
We compare them with that of CN Cha
and clarify what are the differences between them.

\subsection{Fast Novae}

The bottom panel in Figure \ref{light.compar} shows, from left to right, 
the very fast novae V1500 Cyg 1975 and V838 Her 1991,
fast nova V1668 Cyg 1978, and moderately fast nova PW Vul 1984\#1. 
They show a sharp optical peak.   The WD masses of these novae have been
obtained by theoretically reproducing their light curves in the decay phase.
The estimated WD mass is 1.2 $M_\sun$ $(X,Y,Z,Z_{\rm CO},Z_{\rm Ne})=
(0.55,0.30,0.02,0.1,0.03)$ for V1500 Cyg \citep[see Appendix of][]{hac14k}, 
1.35 $M_\sun$ (0.55,0.33,0.02,0.03,0.07) for V838 Her
\citep{kat09v838her}, 0.98 $M_\sun$ $(0.45,0.18,0.02,0.35,0.0)$
for V1668 Cyg \citep{hac16k}, 0.83 $M_\sun$ $(0.55,0.23,0.02,0.2,0.0)$
for PW Vul \citep{hac15k}. 


\subsection{PU Vul: a Slow Nova with a Flat Peak}

The symbiotic nova PU Vul is a unique nova having a long-lasted optical
flat-peak (see Figure \ref{light.compar}a). 
A very stable flat peak lasted as long as eight years.  
The early spectra mimicked those of an F supergiant 
and no indication of strong mass ejection \citep{yam82, kan91a}.
The optical spectrum was absorption-dominated until JD 2,446,000, 
but showed a distinct nebular feature on JD 2,447,000 \citep{iij89,kan91b}.
On JD 2,448,000, the optical and UV spectra showed rich emission lines 
which are typical in the nebular phase \citep{vog92,kan91b,tom91}.
P Cygni line-profiles appeared in the decay phase,   
indicating optically thin mass-ejection from the WD photosphere   
\citep{bel89,vog92,sio93b,nus96}. 

Based on these observational aspects, 
\citet{kat12} presented a light curve model for PU Vul  
with the assumption of hydrostatic envelope, i.e., no optically thick winds. 
They calculated a theoretical light curve that fits with 
the observed UV 1455 \AA~ and optical $V$ data. 
They used a narrow (20\AA\  wide) spectral band centered at 1455~\AA, 
which is known to be emission-line free and can be a representative
of continuum flux in classical novae \citep{cas02}. 

Their model shown in Figure \ref{light.compar} is 
a $0.6~M_\sun$ WD with the optically-thin 
mass-loss rate of $5 \times 10^{-7}~M_\sun$ yr$^{-1}$   
in the decay phase. The chemical composition of the envelope 
is assumed to be $X=0.7$, $Y=0.29$, and $Z=0.01$ because the 
spectra show no indication of C, O, or Ne enrichment. 
The blue line shows the $V$ magnitude of the photospheric 
emission. This model is obtained from the UV light curve fitting. 
The black line represent the total $V$ flux which is the summation 
of the photospheric emission and the emission from optically thin 
plasma surrounding the WD. The origin of the plasma is optically 
thin mass-loss in the decay phase \citep[see][for more detail]{kat12}.

\subsection{Slow novae with Multiple Peaks} 

The slow nova V723 Cas shows an oscillatory behavior 
around $M_V \sim -5$ in the early phase which is settled down to 
a smooth decline in the later phase (Figure \ref{light.compar}b). 
The WD mass in V723 Cas was estimated from a model light curve analysis 
to be 0.5-0.55 $M_\sun$ with the chemical composition of
$(0.55,0.23,0.02,0.2,0.0)$ \citep{hac15k}. 

\citet{kat11drag} presented the light curve model assuming that 
the outburst of V723 Cas began as a hydrostatic envelope without winds
like the PU Vul evolution, but somewhat later changed to a normal evolution 
with smooth light curve decline, i.e., optically thick wind phase.
This model is indicated by the black line in Figure \ref{light.compar}b. 

Such a transition from hydrostatic to winds does not normally occur
because a redgiant-like hydrostatic envelope has a very different 
structure from that of a wind mass-loss envelope. 
\citet{kat11drag} concluded that 
the transition could occur only when the extended nova envelope 
engulfs the companion and its motion in the envelope 
triggered the transition. 
The large energy release owing to the frictional energy 
produces additional luminosity.  We see a relaxation process
in a large spike-like oscillation of the light curve. 

We do not expect such a transition in CN Cha or PU Vul
because they are a very wide binary and 
their nova envelopes/photospheres do not engulf the companion.
  


\begin{figure}
\epsscale{1.15}
\plotone{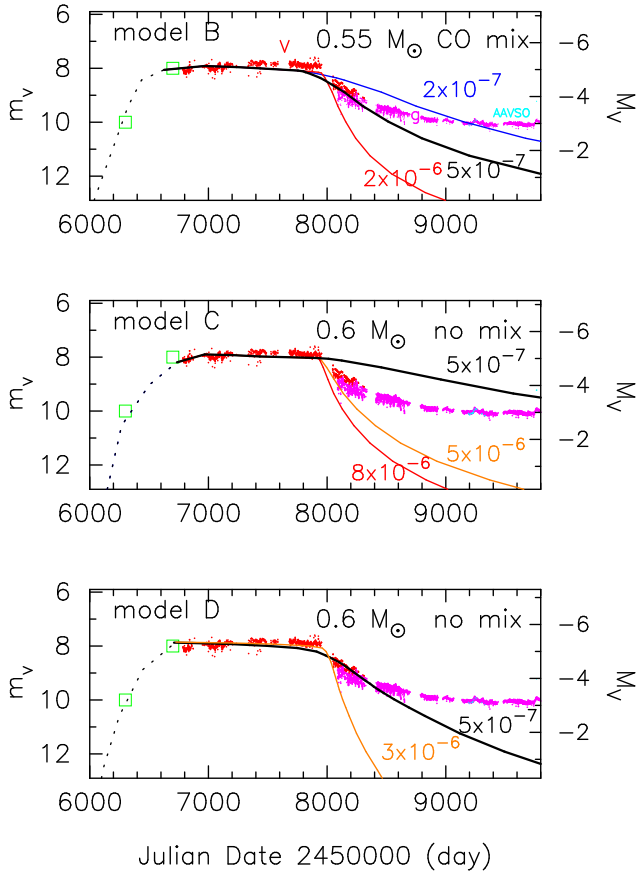}
\caption{
{\bf Top panel:}
Same as those in Figure \ref{m057.CO}(top), but for model B. 
{\bf Middle panel:} Model C. 
{\bf Bottom panel:} Model D. 
\label{light.4}}
\end{figure}

\subsection{CN Cha}



Figure \ref{m057.CO} shows the optical data for CN Cha, 
$V$ (red dots) and $g$ (magenta dots) taken from ASAS-SN 
\citep{sha14, jay19}, and $V$ (cyan blue dots) from AAVSO.
We see that the flat phase variation in CN Cha is 
much smaller than that in V723 Cas, or in 
other similar novae, like HR Del, V5558 Sgr,  
in which optical/$V$ magnitudes change by 2-3 mag 
\citep[Figure 6 in ][]{kat11drag}. 

Lancaster et al.'s spectrum taken on 
JD 2458554.672 (downward arrow labeled ``Sp'' in Figure \ref{light.compar}a) 
shows many narrow emission lines including P Cygni profiles, 
suggesting optically-thin wind mass loss. 
Although we do not find UV flux data in the decay phase of CN Cha,
its resemblance to PU Vul suggests that the $V$ band flux is dominated 
by that of optically thin plasma outside the WD photosphere. 
Our model in Figure \ref{m057.CO} is calculated from the photospheric 
blackbody emission, which does not include the $V$ flux from optically-thin
nebular emission.  Thus, we may take the observed data as the upper
limit for our model $V$ light curve.

\begin{figure}
\epsscale{1.0}
\plotone{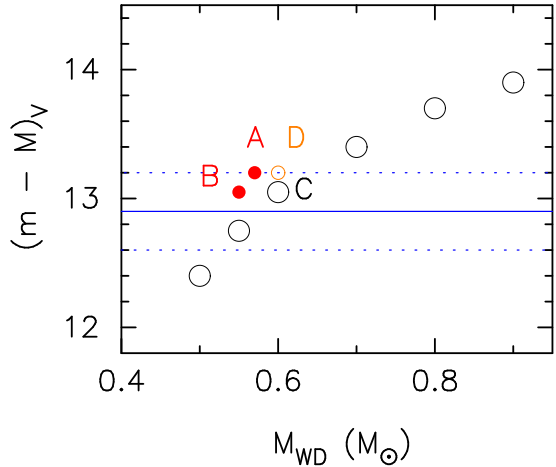}
\caption{
Distance modulus in the $V$ band obtained from our theoretical light curve 
fittings with the flat-peak phase of CN Cha. 
Filled red dot: model A and model B.
Open black circles: models of no CO enrichment including model C. 
Open orange circle: model D.
The horizontal solid/dotted blue lines denote $(m-M)_V=12.9\pm 0.3$.
\label{MvMwd}}
\end{figure}

\section{Model Light Curve Fittings}
\label{section_model}

\subsection{Method}
\label{section_method}

In slow novae, the envelope is almost in hydrostatic balance until
it reaches optical maximum. In the phases of flat optical peak and
after that, the energy generation rate of nuclear burning is balanced with 
the energy loss from the photosphere, i.e., $L_{\rm nuc}= L_{\rm ph}$. 
First, we integrated the hydrostatic equation with 
the diffusion equation from the photosphere 
to the base of envelope.
We use the OPAL opacity \citep{igl93}. 
The model parameters are the WD mass, WD radius (radius at the base of
envelope), and chemical composition of envelope.  
Then, we make a sequence of envelope solutions 
in the order of decreasing mass. 
This sequence is a good approximation of nova evolution 
when no optically thick winds occur.  
The time interval $ \Delta t$ between two successive solutions
is obtained from 
$\Delta t= \Delta M_{\rm env} / (\dot M_{\rm nuc} + \dot M_{\rm wind})$,
where $\Delta M_{\rm env}$ is the difference between the envelope masses
of the two successive solutions, and $\dot M_{\rm wind}$ and 
$\dot M_{\rm nuc}$ are the wind mass-loss rate of optically thin 
winds and hydrogen nuclear burning rate, respectively.  
Here, $\dot M_{\rm nuc}$ is calculated from the envelope structure
and chemical composition of envelope, but we assume the value of
$\dot M_{\rm wind}$ as mentioned below.
This method has been used to follow the supersoft X-ray phase 
of novae after the optically thick wind stops \citep{kat94h,sal05} 
or to follow PU Vul evolution in which the optically thick 
winds does not occur throughout the outburst 
\citep{kat11,kat12}. 
In the flat-peak phase, we have no wind mass loss, 
$\dot M_{\rm wind}=0$.
In the later decay phase of $T_{\rm ph} > 10,000$ K, 
we assume optically thin mass-loss ($\dot M_{\rm wind} \sim $ several 
$\times 10^{-7} ~M_\odot$ yr$^{-1}$)
after \citet{kat11, kat12} in PU Vul. 

In the rising phase, we assume that the envelope is in 
hydrostatic balance, but the energy generation rate is slightly
larger than the thermal equilibrium, i.e., $L_{\rm nuc} > L_{\rm ph}$, 
by a few to several percent. 
We obtained envelope solutions to be consistent with the 
observational data. This part is plotted with a dotted line 
in Figure \ref{m057.CO}(top).

The ignition mass is approximately obtained as the mass 
at $t=0$ when the photospheric temperature goes down to $\log T$ (K)=4.9,
as shown in the middle panel of Figure \ref{m057.CO}.
The envelope mass decreases only by a few percent from $t=0$ to $t=600$ day.

We assume Population II composition for the accreted matter, i.e.,
$Z=0.004$ with/without additional CO enhancement $Z_{\rm CO}$. 
For comparison, we calculated model D ($Z=0.01$) 
as listed in Table \ref{table_model}.


The absolute $V$ magnitudes, $M_V$, for the standard Johnson $V$ bandpass
are calculated from the photospheric temperature, $T_{\rm ph}$,
and luminosity, $L_{\rm ph}$. 
The UVW2 band flux is calculated from the blackbody emission 
between 1120 and 2640 \AA.

\subsection{Light Curve Fitting}

Table \ref{table_model} lists our models. It shows, from left to right, 
the WD mass $M_{\rm WD}$, chemical composition of the envelope
$X$, $Y$, $Z$, $Z_{\rm CO}$, WD radius, ignition mass $M_{\rm ig}$,
distance modulus in the $V$ band $(m-M)_V$,
and assumed optically-thin wind mass-loss rates $\dot {M}_{\rm wind}$. 

The top panel in Figure \ref{m057.CO} shows the light curve fitting
of model A with the observation. 
The model light curve shows a flat optical peak, which is 
the most expanded phase of photosphere,
and its photospheric temperature is as low as 
$T_{\rm ph} < 10,000$ K, as shown in the middle panel. 
We fit our theoretical curve with the observation
to obtain the distance modulus in the $V$ band, i.e., 
$(m-M)_V= m_V$(obs) $- M_V$(model)=13.2. 

The envelope mass gradually decreases because of hydrogen burning 
at a rate of $\dot M_{\rm nuc} \sim 1.7\times 10^{-7}~M_\sun$
yr$^{-1}$.
In the decay phase of $T_{\rm ph} > 10,000$ K,
we assume the optically-thin wind mass-loss as done 
in the PU Vul model by \citet{kat11} and \citet{kat12}.  
For simplicity, we assume a constant mass loss rate ($\dot M_{\rm wind})$. 
The $V$ magnitude decays as $T_{\rm ph}$ increases with time, 
while the photospheric luminosity $L_{\rm ph}$ is almost constant. 
If we assume a larger mass-loss rate, 
the $V$ magnitude decays faster because the envelope mass 
decreases more quickly. 

From the resemblance of the light curve to PU Vul and the presence 
of emission lines that indicate the optically thin mass-loss, we regard 
that the $g$ light curve is strongly 
contaminated with nebular emission 
as often observed in the nebular phase of novae 
(e.g., \citet{kat12} for PU Vul, \citet{hac06kb} for V1668 Cyg).
In other words, the photospheric (continuum) component would 
be much fainter than $g$ light curve. 

We plot three light curves with different wind mass-loss rates 
in the top panel of Figure \ref{m057.CO}. 
The mass loss starts at the point denoted by the open small circle 
in the middle panel. 
The model with $\dot M_{\rm wind}= 1 \times 10^{-7}~M_\odot$ yr$^{-1}$ 
is too slow and rejected. 

Figure \ref{light.4} shows the light curve fittings of 
model B, model C, and model D. 
We obtain the distance modulus in the $V$ band from the flat-peak phase, 
and the lower limit of mass-loss rates from the decay phase. 

It is difficult to further constrain the mass loss rate in CN Cha.  
The bottom panel in Figure \ref{m057.CO} shows
the UV light curves of the three different mass-loss models. 
If we have UV observations in the decay phase, we can determine 
the mass-loss rate by comparing them with the model light curves.

In PU Vul, UV 1455~\AA~ band light curve is obtained with IUE    
that represents temperature evolution in the decay phase. 
\citet{kat11} and \citet{kat12} determined the optically-thin 
wind mass-loss rate to be 
$\dot M_{\rm wind}= (2-5) \times10^{-7}~M_\sun$~yr$^{-1}$ for 
a $\sim 0.6 ~M_\sun$ WD.

The mass loss rate obtained in PU Vul may be a representative of mass loss
rates in the extended nova envelope
for the particular case of no optically thick winds. 
We plot the model of $\dot M_{\rm wind}= 5\times10^{-7}~M_\sun$~yr$^{-1}$ 
by the thick solid black line in Figures \ref{m057.CO} and \ref{light.4}.  
We thus exclude model C.  

For a less massive WD ($M_{\rm WD} \lesssim 0.55~M_\sun$),
the envelope mass is larger while the nuclear burning rate
$\dot M_{\rm nuc}$ is smaller. 
Thus, the evolution timescale is much more longer than those 
in models B and C. 
To reproduce a reasonable light curve for CN Cha,
we need to take a much more larger $\dot M_{\rm wind}$. 
Thus, less massive WDs than those in models B and C are unlikely.

\subsection{Distance Modulus}

Figure \ref{MvMwd} shows the distance modulus in the $V$ band, $(m-M)_V$,
calculated from light curve fitting. 
The photospheric luminosity at the flat-peak is brighter for more massive WDs, 
so that the $(m-M)_V$ becomes larger with the increasing $M_{\rm WD}$. 
The horizontal solid/dotted blue lines indicate $(m- M)_V=12.9\pm 0.3$,  
calculated from a Gaia eDR3 distance, $d=3.05^{+0.19}_{-0.17}$ kpc
\citep{bai21rf}, and scattering in ASAS-SN $V$ data. 
From this plot, we may exclude 
$M_{\rm WD} \lesssim 0.5~M_\sun$ and $M_{\rm WD} > 0.6~M_\sun$. 
Thus, models A and B are reasonable for CN Cha.


For models A, B, and C, we adopt a WD (radius) in a thermal balance
with a relatively large mass-accretion rate 
of several $\times 10^{-8}~M_\sun$ yr$^{-1}$ \citep{kat20}. 
For model D, we take a smaller WD radius for a colder WD for comparison, 
and increase the heavy element enrichment to $Z=0.01$. 
The resultant distance modulus $(m-M)_V$ increases by 0.15 mag, 
because model D is brighter than 
model C mainly because of a smaller WD radius.   
Note that model D is for a Population I star and disfavored 
as a model of CN Cha. 


\section{Discussion}\label{discussion}

We could not accurately obtain/constrain the mass accretion rate because
our static-sequence approach cannot be applied to the early rising phase
of a shell flash.  However, the mass accretion rate corresponding to 
our ignition mass can be found in literature. 

\citet{chen19} presented hydrogen shell flash calculation 
and obtained the ignition mass of 
$M_{\rm ig}=3.3 \times 10^{-5}~M_\sun$ (the recurrence time
$P_{\rm rec}=2,500$ yr)
for a 0.6 $M_\sun$ WD with $Z=10^{-4}$ and CO-rich accretion of 
$\dot M_{\rm acc}=1\times 10^{-8}~M_\sun$ yr$^{-1}$.   
This ignition mass is close to that of model A, which suggests that 
the mass accretion rate of CN Cha is around $10^{-8}~M_\sun$ yr$^{-1}$.  

\citet{chen19} also obtained the ignition mass of 
$M_{\rm ig}=2.2 \times 10^{-4}~M_\sun$ ($P_{\rm rec}=22,000$ yr)
for a $0.6~M_\sun$ WD with no CO enrichment, $Z=10^{-4}$,
and $\dot M_{\rm acc}=1\times 10^{-8}~M_\sun$ yr$^{-1}$.
\citet{kat20} obtained the ignition mass 
to be $M_{\rm ig}=2.0 \times 10^{-4}~M_\sun$ ($P_{\rm rec}=10,000$ yr)
for a $0.6 ~M_\sun$ WD with $Z=0.001$,  no CO enhancement, and 
$\dot M_{\rm acc}=2 \times 10^{-8}~M_\sun$ yr$^{-1}$. 
These values are roughly consistent with model C. 

The above values indicate that the mass accretion rate of CN Cha 
is $\dot M_{\rm acc}\sim 1\times 10^{-8}~M_\sun$ yr$^{-1}$.

\section{Concluding Remarks} \label{conclusion}

CN Cha is a rare nova that has a long-lasted flat-peak. 
Our 0.55-0.57 $M_\sun$ WD models show reasonable $V$ light curve 
fittings with the observation. 
This is the second well-observed flat-peak nova after PU Vul 
that also hosts a less massive WD of $M_{\rm WD}\sim 0.6 ~M_\sun$. 

On the other hand, some slow novae, V723 Cas, HR Del, and V5558 Sgr
have similar WD masses of $M_{\rm WD}\sim 0.55$-$0.6 ~M_\sun$
\citep[e.g.,][]{hac15k}, but they show violent multiple spike-like peaks,
instead of a flat peak. 

We point out that this difference arises from the binary nature: 
a wide binary or not. 
In a close binary like V723 Cas, a nova envelope at/near 
optical maximum extends beyond the binary orbit. In other words, 
the companion main-sequence star moves in the nova envelope.

This idea is supported by the following theoretical implications:
(1) In low mass WDs, there are two possible nova evolutions, which are
represented by an optically-thick wind mass-loss envelope 
and static envelope with no winds \citep{kat09}.
The evolutions of these two envelope solutions do not cross each other
because their envelope structures are very different. 
(2) Transition from static to wind mass-loss solutions could occur, 
if the companion star moves in the envelope 
because the static envelope structure becomes close to 
that of wind solution considering both the centrifugal force
and companion's gravity \citep{kat11drag}.
The observed multiple spike-like peaks are interpreted as
a relaxation process of the transition to wind solutions. 

The main theoretical points of the present work are summarized as follows: 
\begin{enumerate}
\item  Nova outbursts accompany, in most,
strong optically-thick wind mass-loss.
Then the nova evolution is fast, and 
the optical light curve shows a sharp peak
in a white dwarf of mass $M_{\rm WD} \gtrsim 0.55~M_\sun$
\citep[e.g., ][]{hac15k}.

\item When the acceleration is too weak to emit optically thick winds 
\citep{kat09}, the nova envelope evolves very slowly and show 
a flat optical peak like PU Vul and CN Cha.  This slow and static
evolution occurs in a low mass
white dwarf of $M_{\rm WD} \lesssim 0.7 ~M_\sun$ \citep{kat11drag}.

\item In the region between the above two cases 
($0.5 ~M_\sun \lesssim M_{\rm WD} \lesssim 0.7~M_\sun$), 
a nova outburst begins first in a hydrostatic manner, 
and later it could change to an evolution with optically-thick 
wind mass-loss due to perturbation by the companion star in the nova
envelope \citep{kat11drag}.

\item This hypothesis predicts that 
such a transition occurs only in close binaries like V723 Cas  
and does not occur in wide binaries like PU Vul. 
CN Cha is the second example of flat-peak novae in 
wide binaries, and strengthens  
the idea by \citet{kat11drag}. 
\end{enumerate}

Thus, we encourage observations for searching slow novae.
We suggest that a long-lasted flat-peak nova 
appears only in long-period binaries, and not in close binaries.  
Photometric and spectroscopic observations are highly valuable.


\begin{acknowledgments}
We are grateful to the anonymous referee for useful comments,
which improved the manuscript.
We also thank L. Lancaster for discussion on ASAS-SN data 
and T. Jayasinghe for updating ASAS-SN sky patrol data of CN Cha. 
We also thank 
the American Association of Variable Star Observers (AAVSO)
for the archival data of CN Cha.

\end{acknowledgments}

\end{document}